Revtex

\documentstyle[floats,prl,aps]{revtex}
\begin{document}
\renewcommand{\thefootnote}{\fnsymbol{footnote}}
\draft
\title{\large\bf 
  Integrable Kondo impurities 
  in the one-dimensional supersymmetric $U$ model of strongly
  correlated electrons}

\author {   Huan-Qiang Zhou \footnote {E-mail: hqz@maths.uq.edu.au}
,  Xiang-Yu Ge 
\footnote
{E-mail: xg@maths.uq.edu.au}
and Mark D. Gould}

\address{Department of Mathematics,University of Queensland,
		     Brisbane, Qld 4072, Australia}

\maketitle

\vspace{10pt}

\begin{abstract}
Integrable Kondo impurities in the one-dimensional supersymmetric $U$  
model of strongly correlated electrons are studied by means of 
the boundary graded quantum inverse
scattering method.
The boundary $K$ matrices depending on the local magnetic moments of the
impurities 
are presented  as  nontrivial realizations of the reflection equation
algebras in an impurity Hilbert space.
Furthermore, the  model Hamiltonian is diagonalized
and the Bethe ansatz equations are derived. 
It is interesting to note that our model exhibits a free parameter 
in the bulk  Hamiltonian but no free parameter exists on the boundaries.
This is in sharp contrast to the impurity models arising from the
supersymmetric	
$t-J$  and extended Hubbard models where there is no  free parameter
in the bulk but there is a free parameter on each boundary.
\end{abstract}




\def\a{\alpha}
\def\b{\beta}
\def\d{\delta}
\def\e{\epsilon}
\def\g{\gamma}
\def\k{\kappa}
\def\l{\lambda}
\def\o{\omega}
\def\t{\theta}
\def\s{\sigma}
\def\D{\Delta}
\def\L{\Lambda}


\def\beq{\begin{equation}}
\def\eeq{\end{equation}}
\def\bea{\begin{eqnarray}}
\def\eea{\end{eqnarray}}
\def\ba{\begin{array}}
\def\ea{\end{array}}
\def\no{\nonumber}
\def\le{\langle}
\def\re{\rangle}
\def\lt{\left}
\def\rt{\right}

\newcommand{\reff}[1]{eq.~(\ref{#1})}

\vskip.3in

Recently there has been substantial research  devoted to the investigation
of the theory of impurities coupled to  Luttinger liquids. 
Such a problem was first considered by Lee and Toner \cite {LT92}. By using the
perturbative renormalization group theory they found that the 
Kondo temperature crosses from a generic power law dependence on 
the Kondo coupling
constant to an exponential one in the infinite limit.  Afterwards, a  ``poor 
man's'' scaling  procedure was
carried out by Furusaki and Nagaosa \cite {FN94} ,who found
a stable strong coupling fixed point  for both antiferromagnetic
and ferromagnetic cases. On the other hand, boundary conformal field
 theory,first developed by Affleck and Ludwig \cite {AL91} for the
 conventional Kondo problem based on a previous
 work by Nozi{\`e}res \cite {Noz74},
leads us to a classification of critical behaviour for the Kondo
problem in the presence of the electron-electron interactions \cite
{FJ}. It turns
out that there are two types of critical behaviour, i.e., either a local 
Fermi liquid with standard low-temperature thermodynamics or the non-Fermi
liquid observed by Furusaki and Nagaosa \cite {FN94}. However, in order 
to get a full
picture about the critical behaviour
of Kondo impurities coupled to Luttinger liquids, some simple
integrable models, as in the conventional Kondo problem
which allow exact solutions \cite {Wie83,And83}, are desirable.

Several integrable magnetic or nonmagnetic impurity problems
describing impurities embedded
in systems of correlated electrons have so far appeared in the
literature. Among them are  versions of the supersymmetric $t-J$
model with impurities \cite {BEF97,SZ,LF98}. 
Such an idea to incorporate an impurity into a
closed chain  dates back to Andrei and Johannesson \cite {AJ84} (see
also \cite {LS88,ZJ89}). However, the model
thus constructed suffers from the lack of backward scattering and results in
a very complicated Hamiltonian which is diffficult to be justified on
physical grounds. Therefore, as observed by Kane and
Fisher \cite {KF92}, it is advantageous to adopt open boundary
conditions with
the impurities situated at the ends of the chain 
when studying  Kondo impurities coupled to
integrable strongly correlated electron systems\cite {PW97,ZG}. 

In this communication, we 
study integrable Kondo impurities in the one-dimensional
supersymmetric $U$ model of strongly correlated electrons ,
which has been extensively studied
in \cite {Bra95,Bed95,Ram96,Pfa96}. 
Two different non-c-number boundary $K$ matrices are
constructed, which turn out to be quite different from those for
the $t-J$ and the
supersymmetric extended Hubbard models \cite {ZG,ZGG}
, due to the fact that no free parameter exists. 
However,it should be emphasized that our new non-c-number boundary 
$K$ matrices are highly nontrivial, 
in the sense that they can not be factorized into the product of a
c-number boundary $K$ matrix and the corresponding local monodromy
matrices. 
Integrability of the models is established by relating the
Hamiltonians to one parameter families of commuting transfer matrices.
The model is solved by means of the coordinate 
Bethe ansatz method and the Bethe ansatz equations are derived.
It is interesting to note that our model exhibits a free parameter 
in the bulk  Hamiltonian but no free parameter exists on the boundaries.
This is in sharp contrast to the impurity models arising from the
supersymmetric	
$t-J$  and extended Hubbard models where there is no  free parameter
in the bulk but there is a free parameter on each boundary.

Let $c_{j,\s}^\dagger$ and $c_{j,\s}$ denote the creation and
annihilation operators of the conduction electrons with spin $\s$ at
site $j$, which satisfy the anti-commutation relations given by
$\{c_{i,\s}^\dagger, c_{j,\tau}\}=\d_{ij}\d_{\s\tau}$, where 
$i,j=1,2,\cdots,L$ and $\s,\tau=\uparrow,\;\downarrow$. Consider the
Hamiltonian which describes two impurities coupled to the
supersymmetric $U$ open chain
\bea
H&=&-\sum _{j=1}^{L-1} \sum _{\sigma}(c^{\dagger}_{j\sigma}c_{j+1\sigma}+h.c.)
  \exp(-\frac {1}{2}\eta n_{j,-\sigma}-\frac {1}{2}
  \eta n_{j+1,-\sigma})
 +t_p\sum _{j=1}^{L-1} (c^{\dagger}_{j\uparrow}c^{\dagger}_{j\downarrow}
  c_{j+1\downarrow}
  c_{j+1\uparrow}+h.c.)+U \sum ^L_{j=1} n_{j\uparrow}n_{j\downarrow}\no\\
 & & +J_a {\bf S}_a \cdot\sum_{\s,\s'} {\tau}_{\s\s'}
 c^{\dagger}_{1 \s}c_{1 \s'}
 +V_a n_1 +U_a n_{1\uparrow} n_{1\downarrow}
  +J_b {\bf S}_b \cdot \sum_{\s,\s'}{\tau}_{\s\s'} c^{\dagger}_{L \s}c_{L \s'}
  +V_b n_L +U_b n_{L\uparrow} n_{L\downarrow},
       \label{ham}
       \eea
where $J_g,~V_g$ and $U_g~(g=a,b)$ are, respectively,
the Kondo coupling constants,the impurity scalar potentials and the
boundary Hubbard-like interaction constants;
${\bf \tau}\equiv(\tau_x,\tau_y,\tau_z)$ are the
usual Pauli matrices with indexes $|1 \rangle=|\downarrow \rangle$ and $|2 \rangle=
|\uparrow \rangle$;
${\bf S}_{g} (g = a,b)$ are the
local moments with spin-$\frac {1}{2}$ located at the left and right ends of
 the system respectively
and $t_p=\frac {U}{2} =e^{-\eta}-1$; 
$n_{j\s}$ is the number density operator
$n_{j\s}=c_{j\s}^{\dagger}c_{j\s}$,
  $n_j=n_{j\uparrow}+n_{j\downarrow}$.
Below we will establish the quantum integrability of the model (\ref
{ham}) for
the following four choices of the coupling constants:
${\rm Case~ A:}~J_g=2\a+2,~V_g={(\a-1)}/{2},~
	     U_g=-{(\a^2+\a+1)}/{\a};
{\rm Case~ B:}~J_g=-{4(\a+1)}/{[(2\a-1)(2\a+3)]},~
	V_g={3}/{[(2\a-1)(2\a+3)]},~
	U_g={3}/{[\a(2\a-1)(2\a+3)]};
{\rm Case~ C:}~J_a=2\a+2,~V_a={(\a-1)}/{2},~
	     U_a=-{(\a^2+\a+1)}/{\a},~
J_b=-{4(\a+1)}/{[(2\a-1)(2\a+3)]},~
	V_b={3/[(2\a-1)(2\a+3)]},~
	U_b={3}/{[\a(2\a-1)(2\a+3)]};
{\rm Case~ D:}~J_a=-{4(\a+1)}/{[(2\a-1)(2\a+3)]},~
	V_a={3}/{[(2\a-1)(2\a+3)]},~
	U_a={3}/{[\a(2\a-1)(2\a+3)]},
J_b=2\a+2,~V_b={(\a-1)}/{2},~
	     U_b=-{(\a^2+\a+1)}/{\a}.$
Here and hereafter, 
$\alpha = {2}/{U}$.
This is achieved  by showing that the Hamiltonian can be derived from
the graded boundary quantum inverse scattering method. 
Indeed,the Hamiltonian of 
the supersymmetric $U$ model with the periodic boundary conditions
commutes with the transfer matrix, which is the supertrace of the
monodromy matrix $T(u)$,
\beq
T(u) = R_{0L}(u)\cdots R_{01}(u). \label{matrix-t}
\eeq
The explicit form of the quantum R-matrix 
$ R_{0j}(u)$ is given in \cite{Bra95}.
Here $u$ is the spectral parameter, 
and the subscript $0$ denotes the auxiliary superspace $V={\bf C}^{2,2}$.
It should be noted that the supertrace
is carried out for the auxiliary superspace $V$.
The elements of the supermatrix $T(u)$ are the generators
of an associative superalgebra ${\cal A}$ defined by the relations
\beq
R_{12}(u_1-u_2) \stackrel {1}{T}(u_1) \stackrel {2}{T}(u_2) =
   \stackrel {2}{T}(u_2) \stackrel {1}{T}(u_1)R_{12}(u_1-u_2),\label{rtt-ttr} 
\eeq
where $\stackrel {1}{X} \equiv  X \otimes 1,~
\stackrel {2}{X} \equiv  1 \otimes X$
for any supermatrix $ X \in End(V) $. For later use, we list some useful
properties enjoyed by the R-matrix:
(i) Unitarity:   $  R_{12}(u)R_{21}(-u) = 1$ and (ii)
 Crossing-unitarity:  $  R^{st_2}_{12}(-u+2)R^{st_2}_{21}(u) =
         \tilde {\rho }(u)$
with $\tilde \rho (u)$ being a scalar function,
$\tilde \rho (u) =u^2(2-u)^2/[(2+2\alpha-u)^2(2\alpha+u)^2]$.

In order to describe integrable electronic models on  open
chains, we introduce two associative
superalgebras  ${\cal T}_-$ and ${\cal T}_+$ defined by the $R$-matrix
$R(u_1-u_2)$ and the relations
\beq
R_{12}(u_1-u_2)\stackrel {1}{\cal T}_-(u_1) R_{21}(u_1+u_2)
  \stackrel {2}{\cal T}_-(u_2)
=  \stackrel {2}{\cal T}_-(u_2) R_{12}(u_1+u_2)
  \stackrel {1}{\cal T}_-(u_1) R_{21}(u_1-u_2)  
  \label{reflection1},
\eeq
\bea
&&R_{21}^{st_1 ist_2}(-u_1+u_2)\stackrel {1}{{\cal T}^{st_1}_+}
  (u_1) R_{12}(-u_1-u_2+2)
  \stackrel {2}{{\cal T}^{ist_2}_+}(u_2)\no\\
&&~~~~~~~~~~~~~~~~~~~~~=\stackrel {2}{{\cal T}^{ist_2}_+}(u_2) R_{21}(-u_1-u_2+2)
  \stackrel {1}{{\cal T}^{st_1}_+}(u_1) R_{12}^{st_1 ist_2}(-u_1+u_2)
  \label{reflection2}
\eea
respectively. Here the supertransposition $st_{\mu}~(\mu =1,2)$ 
is only carried out in the
$\mu$-th componant of the superspace $V \otimes V$, whereas $ist_{\mu}$ denotes
the inverse operation of  $st_{\mu}$. By modifying Sklyanin's 
arguments \cite{Skl88}, one
may show that the quantities $\tau(u)$ given by
$\tau (u) = str ({\cal T}_+(u){\cal T}_-(u))$
constitute a commutative family, i.e.,
        $[\tau (u_1),\tau (u_2)] = 0$ \cite {Zhou97,BRA97} . 

One can obtain a class of realizations of the superalgebras ${\cal T}_+$  and
${\cal T}_-$  by choosing  ${\cal T}_{\pm}(u)$ to be of the form
\beq
{\cal T}_-(u) = T_-(u) \tilde {\cal T}_-(u) T^{-1}_-(-u),~~~~~~ 
{\cal T}^{st}_+(u) = T^{st}_+(u) \tilde {\cal T}^{st}_+(u) 
  \lt(T^{-1}_+(-u)\rt)^{st}\label{t-,t+} 
\eeq
with
\beq
T_-(u) = R_{0M}(u) \cdots R_{01}(u),~~~~
T_+(u) = R_{0L}(u) \cdots R_{0,M+1}(u),~~~~ 
\tilde {\cal T}_{\pm}(u) = K_{\pm}(u),
\eeq
where $K_{\pm}(u)$, called boundary K-matrices, 
are representations of  ${\cal T}_{\pm}  $ in 
some representation superspace. Although many attempts have been made to
find c-number boundary $K$ matrices, which may be referred to as the
fundamental representation, it is no doubt very intereting to search for
non-c-number $K$ matrices, arising as representations in some Hilbert
spaces, which may be interpreted as impurity Hilbert spaces \cite{ZG}.

We now solve (\ref{reflection1}) and (\ref{reflection2}) 
for $K_-(u)$ and $K_+(u)$. Quite interestingly,
for the supersymmetric $U$ model \cite{Bra95},there are two different
non-c-number boundary $K$ matrices.
One is
\beq
K^I_-(u)=   \left ( \begin {array}
{cccc}
1&0&0&0\\
0&A^I_-(u)&B^I_-(u)&0\\
0&C^I_-(u)&D^I_-(u)&0\\
0&0&0&1
\end {array} \right ),\label{K1}
\eeq
where
$
A^I_-(u)=-(u^2-2u+4-u {\bf S}^z_a)/Z^I_-,~
B^I_-(u)=2u {\bf S}^-_a/Z^I_-,~
C^I_-(u)=2u {\bf S}^+_a/Z^I_-,~
D^I_-(u)=-(u^2-2u+4+u {\bf S}^z_a)/Z^I_-, ~Z^I_- \equiv (u-2)(u+2),~
$
and the other takes the form,
\beq
K^{II}_-(u)=   \left ( \begin {array}
{cccc}
1&0&0&0\\
0&A^{II}_-(u)&B^{II}_-(u)&0\\
0&C^{II}_-(u)&D^{II}_-(u)&0\\
0&0&0&F_-(u)
\end {array} \right ),\label{K2}
\eeq
with
$A^{II}_-(u)=- {(u^2-2u-4\a^2-4\a+3-u {\bf S}^z_a)}/Z^{II}_-,~
B^{II}_-(u)= {2u {\bf S}^-_a}/Z^{II}_-,~
C^{II}_-(u)= {2u {\bf S}^+_a}/Z^{II}_-,~
D^{II}_-(u)=-{(u^2-2u-4\a^2-4\a+3+u {\bf S}^z_a)}/Z^{II}_-,~
F_-(u)={((u+2\a-1)(u+2\a+3))}/Z^{II}_-, ~Z^{II}_- \equiv (u-2\a+1)(u-2\a-3).
$
Here ${\bf S}^{\pm}={\bf S}^x \pm
i{\bf S}^y$.
The matrix $K_+(u)$ can be obtained from the isomorphism of the
superalgebras  ${\cal T}_-  $ and ${\cal T}_+  $. Indeed, given a solution
$K_-(u) $ of the equation (\ref{reflection1}), then $K_+(u)$ defined by
 $K_+^{st}(u) =   K_-(-u+\frac {1}{2})$
is a solution of the equation (\ref{reflection2}). 
The proof follows from some algebraic computations 
by making use
of the properties of the R-matrix\cite {BRA97}.
Therefore, one may choose the boundary matrix $K_+(u)$ as 
\beq
K^I_+(u)=   \left ( \begin {array}
{cccc}
1&0&0&0\\
0&A^I_+(u)&B^I_+(u)&0\\
0&C^I_+(u)&D^I_+(u)&0\\
0&0&0&F_+(u) 
\end {array} \right ),
\eeq
where
$
A^I_+(u)=- {(u^2-4\a^2-4\a+2 -(u-1) {\bf S}^z_b)}/Z^I_+,~
B^I_+(u)={2(u-1){\bf S}^-_b}/Z^I_+,~
C^I_+(u)={2(u-1){\bf S}^+_b}/Z^I_+,~
D^I_+(u)=- {(u^2-4\a^2-4\a+2 +(u-1) {\bf S}^z_b)}/Z^I_+,~
F_+(u)={((u-2\a)(u-2\a-4))}/Z^I_+,~ Z^I_+ \equiv {(u+2\a+2)(u+2\a-2)},
$
and
\beq
K^{II}_+(u)=   \left ( \begin {array}
{cccc}
1&0&0&0\\
0&A^{II}_+(u)&B^{II}_+(u)&0\\
0&C^{II}_+(u)&D^{II}_+(u)&0\\
0&0&0&1 
\end {array} \right ),
\eeq
where
$
A^{II}_+(u)=-{(u^2+3 -(u-1) {\bf S}^z_b)}/Z^{II}_+,~
B^{II}_+(u)={2(u-1){\bf S}^-_b}/Z^{II}_+,~
C^{II}_+(u)={2(u-1){\bf S}^+_b}/Z^{II}_+,~
D^{II}_+(u)=-{(u^2+3 +(u-1) {\bf S}^z_b)}/Z^{II}_+,~Z^{II}_+ \equiv
(u+1)(u-3).
$

As usual,the boundary transfer matrix $\tau(u)$ may be rewritten as
\beq
\tau(u)=str[K_+(u)T(u)K_-(u)T^{-1}(-u)],
  \eeq
  Since $K_\pm(u)$  can be taken as $K^I_\pm(u)$ or $K^{II}_\pm(u)$,
  respectively, we have four possible choices of boundary transfer
  matrices,
  which reflects the fact that
  the boundary conditions on the left end and on the right end of
  the open lattice chain are independent.
  Then it can be shown \cite{Zhou97,BRA97} that Hamiltonians corresponding to
  all four 
  chocices can be embedded into the above four boundary transfer
  matrices,
  respectively.Indeed, the 
Hamiltonian (\ref{ham}) is related to the transfer matrix
$\tau (u)$ (up to an unimportant additive chemical potential term)
\bea
H^R \equiv -\frac {U}{2(U+2)}H=\frac {\tau'' (0)}{4(V+2W)}&=&
  \sum _{j=1}^{L-1} H^R_{j,j+1} + \frac {1}{2} \stackrel {1}{K'}_-(0)
+\frac {1}{2(V+2W)}\lt[str_0\lt(\stackrel {0}{K}_+(0)G_{L0}\rt)\rt.\no\\
& &\lt.+2\,str_0\lt(\stackrel {0}{K'}_+(0)H_{L0}^R\rt)+
  str_0\lt(\stackrel {0}{K}_+(0)\lt(H^R_{L0}\rt)^2\rt)\rt],\label{derived-h}
\eea
where 
$ V=str_0 K'_+(0),
~W=str_0 \lt(\stackrel {0}{K}_+(0) H_{L0}^R\rt),~
H^R_{i,j}=P_{i,j}R'_{i,j}(0),
~G_{i,j}=P_{i,j}R''_{i,j}(0),
$with $P_{i,j}$ being the graded permutation operator acting on the $i$-th
and $j$-th quantum spaces. (\ref{derived-h}) implies that the boundary
supersymmetric $U$ model admits
an infinite number
of conserved currents which are in involution with each other, thus
assuring the integrability. It should be emphasized that 
Hamiltonian (\ref{ham}) appears as the second derivative of the transfer matrix
$\tau (u)$ with respect to the spectral parameter $u$ at $u=0$. This
is due to the fact that the supertrace of $K_+(0)$ equals zero.
As pointed out in Ref. \cite {BRA97}, the reason for the zero supertrace of $K_+(0)$
is related to the fact that the quantum space is a
4-dimensional {\em typical} irreducible representation of $gl(2|1)$. 
A similar situation also occurs in the Hubbard-like models \cite{Zhou96}.

The Hamiltonian (\ref {ham}) may be diagonalized by means of the coordinate
Bethe Ansatz method.
The Bethe ansatz equations are 
\bea
(\frac {\theta _j-\frac {i}{2}}
 {\theta _j+\frac {i}{2}})^{2L}
\prod _{g=a,b}\frac{\theta _j-\theta _g+ic}{\theta _j+\theta _g-ic}
&=&\prod ^M_{\beta =1}\frac {\theta _j-\lambda_\beta +i\frac {c}{2}}
  {\theta _j-\lambda_\beta -i\frac {c}{2}}
  \cdot\frac {\theta _j+\lambda_\beta +i\frac {c}{2}}
  {\theta _j+\lambda_\beta -i\frac {c}{2}},\no\\
\prod _{g=a,b}\frac{(\lambda_\a+\frac{ic}{2})^2-{\theta_g} ^2}
{(\lambda_\a-\frac{ic}{2})^2-{\theta_g }^2}
\prod ^N_{j =1}\frac {\lambda _{\alpha}-\theta _j +i\frac {c}{2}}
  {\lambda _{\alpha}-\theta_j -i\frac {c}{2}}
  \cdot\frac {\lambda_{\alpha}+\theta_j +i\frac {c}{2}}
  {\lambda_{\alpha}+\theta _j -i\frac {c}{2}}
&=&
  \prod ^M_{\stackrel {\beta =1}{\beta \neq \alpha}}\frac
  {\lambda _{\alpha}-\lambda _{\beta} +ic}
  {\lambda _{\alpha}-\lambda_{\beta} -ic}
  \cdot\frac {\lambda_{\alpha}+\lambda_{\beta} +ic}
  {\lambda_{\alpha}+\lambda _{\beta} -ic},\label {bet}
\eea
where $c=e^\eta-1$,
the charge rapidities 
$\t_j\equiv \t(k_j)$ 
are related to the single-particle
quasi-momenta $k_j$ by $\theta (k)=\frac {1}{2} \tan (\frac {k}{2})$
\cite{Bed95}, and
 $\theta _a,
~\theta _b$ take the following form for the four choices :
${\rm Case~ A:}~ \theta _a=-\frac{i}{2},
~ \theta _b=-\frac{i}{2} ;~
{\rm Case~ B:}~ \theta _a=\frac{i}{U+2},
~ \theta _b=\frac{i}{U+2};
{\rm Case~ C:}~ \theta _a=-\frac{i}{2},
~ \theta _b=\frac{i}{U+2};
{\rm Case~ D:}~ \theta _a=\frac{i}{U+2},
~ \theta _b=-\frac{i}{2}.$
The corresponding energy eigenvalue $E$ of the model is given by
$E=-2\sum ^N_{j=1}\cos k_j$, where we have dropped an additive constant.

In conclusion, we have studied integrable Kondo impurities coupled with
the one-dimensional supersymmetric $U$ open chain. The quantum integrability
follows from the fact
that the model Hamiltonian may be embbeded into
a one-parameter family of commuting transfer matrices. Moreover, the Bethe
Ansatz equations are derived by means of the coordinate Bethe ansatz
approach. 
It is quite interesting to note that in the boundary $K$ matrices (\ref
{K1}) and
(\ref {K2}),no free parameter is available ,in contrast to the $t-J$ and 
the supersymmetric extended Hubbard
models \cite {ZG,ZGG}. Further,it is desirable to investigate  the
thermodynamic equilibrium properties of the model, based on the Bethe
ansatz equations (\ref {bet}). The details are deferred to another publication.

This work is supported by OPRS and UQPRS.


\end{document}